\documentclass[review]{elsarticle}

\usepackage{lineno,hyperref}
\modulolinenumbers[5]

\journal{Pattern Recognition}

\usepackage{bm}
\usepackage{amsmath}
\usepackage{rotating}
\usepackage{color, colortbl}

\def\x{\bm{x}}
\def\t{\bm{t}}
\def\o{\bm{o}}
\def\y{\bm{y}}

\def\Y{\bm{Y}}

\def\W{\bm{W}}
\def\u{\bm{u}}
\def\s{\bm{s}}
\def\W{\bm{W}}
\def\x{\bm{x}}

\def\Y{\bm{Y}}

\def\W{\bm{W}}

\definecolor{c1}{rgb}{0.88,1,1}
\definecolor{c2}{rgb}{1,0.88,1}
\definecolor{c3}{rgb}{1,1,0.88}
\definecolor{c4}{rgb}{0.88,0.88,0.88}

\begin{document}

\begin{frontmatter}

\title{MIXCAPS: A Capsule Network-based Mixture of Experts for Lung Nodule Malignancy  Prediction}
\author[a]{Parnian Afshar}
\author[a]{Farnoosh Naderkhani}
\author[b]{Anastasia Oikonomou}
\author[c]{Moezedin Javad Rafiee}
\author[a]{Arash Mohammadi\corref{cor1}}
\author[d]{Konstantinos N.  Plataniotis}
\cortext[cor1]{Corresponding author:
  Tel.: +1 (514) 848-2424 ext. 2712; This work was partially supported by the Natural Sciences and Engineering Research Council (NSERC) of Canada through the NSERC Discovery
Grant RGPIN-2016-04988.}
\ead{arash.mohammadi@concordia.ca}

\address[a]{Concordia Institute for Information Systems Engineering, Concordia University, Montreal, QC, Canada}
\address[b]{Department of Medical Imaging, Sunnybrook Health Sciences Centre, University of Toronto, Canada}
\address[c]{Affiliation 1: Department of Medicine and Diagnostic Radiology, McGill University Health Center- Research Institute. Affiliation 2: Babak Imaging Center, Tehran, Iran}
\address[d]{Department of Electrical and Computer Engineering, University of Toronto, Toronto, ON, Canada}
\begin{abstract}
Lung diseases including infections such as Pneumonia,  Tuberculosis, and novel Coronavirus  (COVID-19), together with Lung Cancer are significantly widespread and are, typically, considered life threatening. In particular, lung cancer is among the most common and deadliest cancers with a low 5-year survival rate. Timely diagnosis of lung cancer is, therefore, of paramount importance as it can save countless lives. In this regard, Computed Tomography (CT) scan is widely used for early detection of lung cancer, where human judgment is currently considered as the gold standard approach. Recently, there has been a surge of interest on development of automatic solutions via radiomics, as human-centered diagnosis is subject to inter-observer variability and is highly burdensome. Hand-crafted radiomics, serving as a radiologist assistant, requires fine annotations and pre-defined features. Deep learning radiomics solutions, however, have the promise of extracting the most useful features on their own in an end-to-end fashion without having access to the annotated boundaries. Among different deep learning models, Capsule Networks are proposed to overcome shortcomings of the  Convolutional Neural Networks (CNNs) such as their inability to recognize detailed spatial relations. Capsule networks have so far shown satisfying performance in medical imaging problems. Capitalizing on their success, in this study, we propose a novel capsule network-based mixture of experts, referred to as the MIXCAPS. The proposed MIXCAPS architecture takes advantage of not only the capsule network's capabilities to handle small datasets, but also automatically splitting dataset through a convolutional gating network. MIXCAPS enables capsule network experts to specialize on different subsets of the data. Our results show that MIXCAPS outperforms a single capsule network and a mixture of CNNs, with an accuracy of $92.88\%$, sensitivity of $93.2\%$, specificity of $92.3\%$ and area under the curve of $0.963$. Our experiments also show that there is a relation between the gate outputs and a couple of hand-crafted features, illustrating explainable nature of the proposed MIXCAPS. To further evaluate generalization capabilities of the proposed MIXCAPS architecture, additional experiments on a brain tumor dataset are performed showing potentials of MIXCAPS for detection of tumors related to other organs.
\end{abstract}
\begin{keyword}
Tumor type classification \sep Capsule network \sep Mixture of experts
\end{keyword}

\end{frontmatter}

\section{Introduction}\label{sec:Introduction}
Lung cancer, according to recent statistics~\cite{Bray:2018}, is associated with the highest mortality rate, among all different cancer types, and is considered as one of the top three cancers, in terms of incidence. The combined 5-year survival for lung cancer is still low~\cite{Siegel:2016}, at 18\%, because the majority of patients are diagnosed at advanced stages~\cite{Xie:2019}. What makes the early diagnosis of lung cancer significantly challenging is the lack of sufficient visible warning symptoms and signs in early stages of the disease. Computed Tomography (CT) scan~\cite{Aberle:2011} is by far one of the most advanced and effective techniques used for lung cancer diagnosis. However, even the CT scans may not reveal convincing signs that can contribute to early diagnosis of lung cancer. In other words, Imaging features of nodule such as size, shape, and attenuation that play an important role in identifying the cancer may not be immediately accessible to the unaided eye~\cite{Causey:2018}. More importantly, human-centered diagnosis is subject to inter-observer variability, meaning that radiologists can have different judgments, depending on their previous experience. Finally, investigating the test results and coming into an inclusive decision can be extremely time-consuming and burdensome~\cite{Zhang:2017}.

Radiomics analysis~\cite{Aerts:2014,Oikonomou:2018,Afshar:2019}, referring to the extraction of several quantitative and semi-quantitative features from the medical images, is one of the most successful approaches towards automatizing the cancer diagnosis/prediction process~\cite{Lambin:2012}. Features extracted in the radiomics analysis are aimed at capturing different properties of the nodules, such as their shape and texture. Such features have shown association with the nodule malignancy, its stage, and even the patient's survival time. Radiomics is often categorized in two groups of hand-crafted~\cite{Chen:2018,Parmar:2015,Coroller:2016,Huynh:2016} and deep learning-based. The former category involves extraction of a set of pre-defined features that are further processed and analyzed by a statistical or Machine Learning (ML) model. Despite showing satisfactory results in different tasks~\cite{Gillies:2015, Oikonomou2:2019}, hand-crafted radiomics is limited to the features defined by the radiologists and as such there is no guarantee that the features contribute to the problem at hand. Furthermore, since  hand-crafted Radiomics features are extracted from the annotated Region of Interest (ROI), they are still subject to inter-observer variability, and besides being time-consuming, their performance highly depends on the accuracy of the provided annotations~\cite{Yip:2016}. In other words, extra effort is required to enhance the annotations and select  features that are more descriptive and robust~\cite{Park:2019}.

Deep learning-based radiomics~\cite{Li:2017,Oakden:2017,Cha:2017}, proposed to overcome the shortcomings of its hand-crafted counterparts, does not require a pre-knowledge about the types of features to be utilized. In other words, deep learning-based techniques are  capable of extracting features that can best contribute to the problem at hand in an end-to-end fashion. Furthermore, deep learning-based radiomics does not need to be fed with the annotated ROI, which has the promise of reducing the effect of inter-observer variability as well as the burden of segmenting the images. Among different deep learning techniques, Convolutional Neural Networks (CNNs) are more popular within the field of radiomics~\cite{Kumar:2017}, due to their ability to efficiently process and learn meaningful features from medical images~\cite{Krizhevsky:2012}. Performance of the CNNs, however,  partly depends on the size of the available dataset~\cite{Yamashita:2018}. More specifically, CNNs, typically, fail to determine the spatial relations between the image instances and identify  rotation or transformation of an object. As such, CNNs need to be fed with a large dataset containing all the possible transformations of the objects. Large datasets are, however, not typically available in medical imaging in particular for lung cancer malignancy prediction.

Capsule networks~\cite{Sabour:2017}, also referred to as the CapsNets, are developed aiming at overcoming the aforementioned drawbacks of the CNNs. CapsNets use capsules, instead of using individual neurons, to represent imaging instances. CapsNets, therefore, can identify the spatial relations via their ``Routing by Agreement" process, through which capsules try to come to a mutual agreement about the existence of the objects. In particular, CapsNet's ability to handle transformations is further investigated in Reference~\cite{LaLondea:2020} for medical image segmentation. In our recent studies~\cite{Afshar:2020, Afshar:2018, Afshar:2019ICASSP}, we showed superior performance and capabilities of CapsNets for tumor type classification.

Capitalizing on the success of the CapsNets, in this study we propose a new framework, referred to as the Mixture of Capsule networks (MIXCAPS), for the task of lung nodule malignancy prediction. The proposed MIXCAPS framework is a ``Mixture of Experts'' type model~\cite{Arash:2015, Jacobs:1991,  Rasti:2017,Guo:2015}, which has the potential to noticeably improve the classification accuracy by integrating/coupling several experts (individual CapsNets in the context of the proposed MIXCAPS). To be more precise, mixture of experts solves the classification problems by splitting the dataset into similar samples, and each expert specializes in classifying similar instances.
To the best of our knowledge, the proposed MIXCAPS is the first CapsNet-based mixture of experts framework. The MIXCAPS model benefits from the following three important properties: (i) The embedded capsule network is capable of classifying the lung nodules without requiring availability of a large dataset; (ii) The mixture of experts approach enables each CapsNet within the MIXCAPS architecture to focus on a specific subset of the nodules, therefore, improving the overall classification performance of the model, and; (iii) As shown in our experiments, MIXCAPS is not restricted to the task of lung nodule malignancy prediction. In fact, it can be easily generalized to the prediction of other tumor types such as brain cancer. The following summarizes our contributions:
\begin{itemize}
\item CapsNets are utilized, for the first time, as individual experts within a mixture of experts framework.
\item A new and modified CapsNet loss function (margin loss) is developed to reflect the loss associated with the experts and gating models.
\item Output of the gating model is investigated for potential correlations with nodule hand-crafted features to improve the potential interpretability of the proposed MIXCAPS.
\item Generalizability of the proposed MIXCAPS is illustrated via extension and evaluation based on a separate dataset associated with a different prediction task other than the one initially used to design the framework.
\end{itemize}
%
The rest of this paper is organized as follows: First, in Section~\ref{sec:Related} the previous studies on lung nodule malignancy prediction is briefly investigated. In Section~\ref{sec:Method}, the dataset and the pre-processing steps are described, along with the proposed MIXCAPS. Results and discussions are presented in Section~\ref{sec:Result}. Finally, Section~\ref{Sec:Conclusion} concludes the paper.

\section{Related Works}\label{sec:Related}
Generally speaking, most of the studies based on hand-crafted radiomics follow a pre-defined set of steps~\cite{Aerts:2014,Oikonomou:2018,Afshar:2019}:
\begin{enumerate}
\item[(i)] The first step is to pre-process the images and segment the nodule;
\item[(ii)] The second and the main step is to extract hundreds of features from the segmented nodule. These features mostly fall into three categories of intensity-based, shape-based, and texture-based features. The former category captures basic properties of the nodule related to its histogram. While shape-based features quantify shape-related properties such as area, diameter, and volume, texture-based ones capture the heterogeneity of the nodule texture;
\item[(iii)] In the third step of the hand-crafted radiomics analysis, feature reduction techniques are utilized to select the most relevant and robust features;
\item [(iv)] In the final step, extracted features are fed to a statistical or machine learning tool to calculate the desired outcome.
\end{enumerate}
For example, the study performed by authors in Reference~\cite{Mao:2019} is a recent implementation of the above mentioned steps for extracting hand-crafted radiomics for lung nodule malignancy prediction. In this study, a total of $385$ features is extracted from the annotated nodules. Consequently, based on a correlation analysis, the non-redundant features are selected and fed to a regression model to output the malignancy probability.

The limitations of the hand-crafted radiomics, including its dependence on the annotated region, have caused a surge of interest in deep learning-based radiomics, especially using CNNs~\cite{Carvalho:2018,Liu:2018}. CNNs are powerful models for analyzing images and extracting features that best contribute to the problem at hand, through trainable filters. Furthermore, filters share weights across the input, which significantly reduces the computational cost, compared to a fully-connected network. CNNs have been recently used for the problem of lung nodule malignancy prediction. While some studies~\cite{Sun:2017,Wang:2017} have proposed to adopt previously developed CNN architectures for the radiomics analysis, others~\cite{Kumar:2017,Fu:2017} have designed and optimized their own specific CNN-based models. Although showing satisfying results, most of these studies had to use a data augmentation or transfer learning strategy to compensate for the lack of large datasets specifically for the problem of lung nodule malignancy prediction. These strategies, however, are associated with more computational costs. Furthermore, there is still no comprehensive study on the effectiveness of these strategies for the nodule malignancy prediction. Capsule network (CapsNet), briefly described in the following section, is an alternative and attractive modeling paradigm to address the aforementioned issues, i.e., accounting for more variations in the input, without resorting to  heavy data augmentation.

\subsection{Capsule Networks}\label{sec:caps}
\begin{figure}[th]
\centering
\includegraphics[scale=.55]{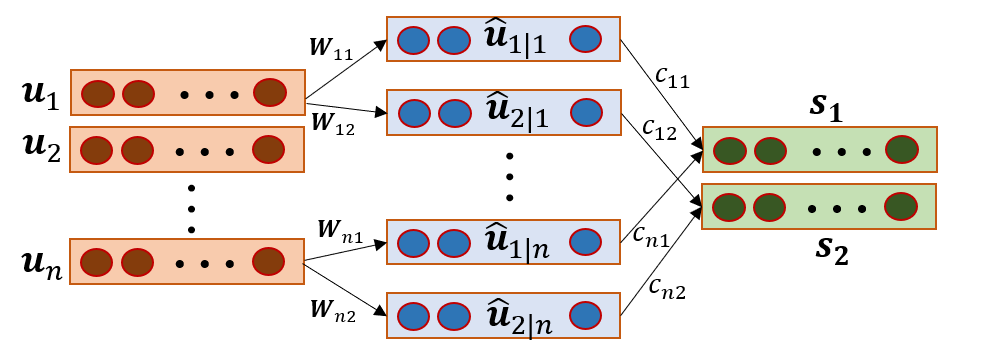}
\caption{Routing by agreement. For the sake of simplicity number of output capsules is set to two.}\label{fig:capsule}
\end{figure}
Capsule networks are constructed based on capsules, as their main building blocks. A capsule being represented by a vector consists of several neurons representing, collectively, a specific object at a specific location. While neurons capture the instantiation parameters of the object, the length of a capsule determines the existence probability of that object. The most important property of a capsule network, distinguishing it from CNNs, is its routing by agreement process. Generally speaking, each Capsule $i$, having the instantiation parameter vector $\u_i$, in a lower layer tries to predict the output of the capsules in the next layer, through a trainable weight matrix $\W_{ij}$ given by
\begin{equation}
\hat{\u}_{j|i}=\W_{ij}\u_i\label{eq:pred},
\end{equation}
where $\hat{\u}_{j|i}$ denotes the prediction for parent Capsule $j$. Through the routing by agreement process, the predictions are evaluated in terms of their similarity to the actual outputs. More weight is then given to the successful predictions, before calculating the final output $\s_j$ for the capsule $j$, as follows
\begin{eqnarray}
 a_{ij} &=& \s_j.\hat{\u}_{j|i},\label{eq:rout}\\
b_{ij} &=& b_{ij}+a_{ij},\\
c_{ij} &=& \frac{\exp(b_{ij})}{\sum_k \exp(b_{ik})},\label{eq:score}\\
\text{ and  }\s_j &=& \sum_ic_{ij}\hat{\u}_{j|i},\label{eq:sj}
\end{eqnarray}
where $a_{ij}$ shows the agreement between actual output $\s_j$ and its prediction $\hat{\u}_{j|i}$, and $c_{ij}$ denotes the score assigned to the prediction based on the obtained agreement. The routing by agreement process, summarized in Fig.~\ref{fig:capsule}, enables capsule the network to recognize spatial information between image instances.

Tumor classification based on capsule networks has been investigated in several recent studies, leading to increased performance when compared to CNNs. Lung tumor malignancy prediction is considered in Reference~\cite{Afshar:2020}, where a multi-scale framework is proposed, outperforming single-scale and multi-scale CNNs. Classifying tumors related to other organs, such as brain, using capsule networks, has also been investigated in several studies~\cite{Afshar:2018, Afshar:2019ICASSP,Adu:2019,Cheng:2019}, leading to satisfying performance. The paper makes a unique contribution in this field by introducing a novel CapsNet architecture based on ``Mixture of Experts'', which is briefly described below.

\subsection{Mixture of Experts}
\begin{figure}
\centering
\includegraphics[scale=.5]{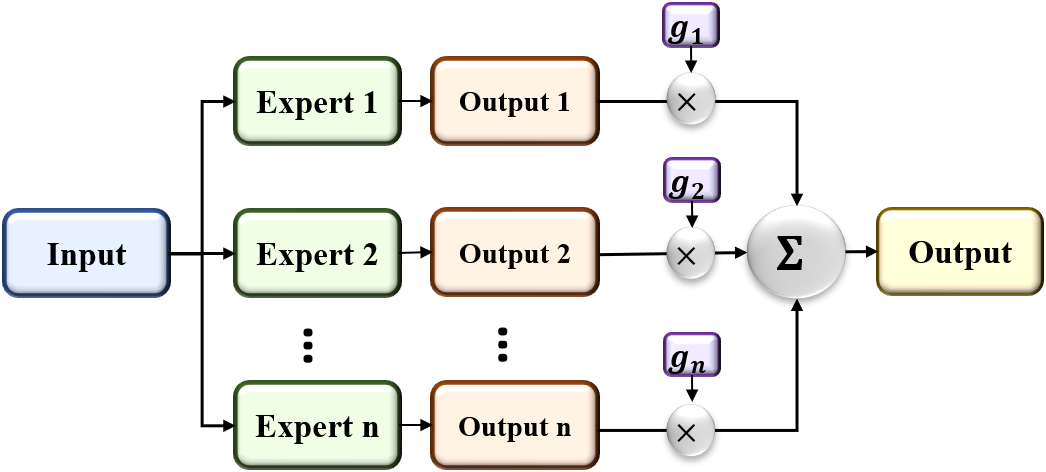}
\caption{General framework of a mixture of experts approach.}\label{fig:moe}
\end{figure}

Mixture of experts (MoE)~\cite{Jacobs:1991} refers to adopting several experts, each of which is specialized on a subset of the data, to collectively perform the final prediction task. As shown in Fig.~\ref{fig:moe}, experts are separately fed with the input data and the final output is a weighted average of all the predictions coming from all the $N$ active  experts. The weight $g_i$ assigned to Expert $i$ can be either a pre-determined value, or a trainable one. One simple example of the former case is averaging over all the experts' predictions~\cite{Guo:2015}. However, more sophisticated approaches such as soft clustering of the input may also be adopted. In the latter case, weights may be trained at the same time with the experts. One other approach to use trainable gating weights is to concatenate  the feature vectors obtained from the individual experts and feed the resulting vector to an external gating model to make the final decision.

The MoE concept has been widely used in medical imaging. The simple averaging scenario is investigated in References~\cite{Maji:2016} and \cite{Chen:2016} for retinal vessel detection from fundus images and breast cancer detection from histology images, respectively. Trainable gating weights are studied in Reference~\cite{Wang:2014}, where hand-crafted and CNN-based features are combined to detect breast cancer from pathology images. The scenario where gating weights are trained at the same time with the experts is investigated in Reference~\cite{Rasti:2017} for breast cancer diagnosis. In particular, CNN experts are combined using weights coming from an external gating network. The gating network itself is a CNN, taking the same inputs as the experts, and outputting the probability of each expert being responsible for each particular input. Our proposed MIXCAPS, which is based on the same gating scenario as Reference ~\cite{Rasti:2017}, is explained in the next section, along with its incorporated data pre-processing approach.

\section{The Proposed MIXCAPS Framework}\label{sec:Method}
\begin{sidewaysfigure}
\centering
\includegraphics[scale=.55]{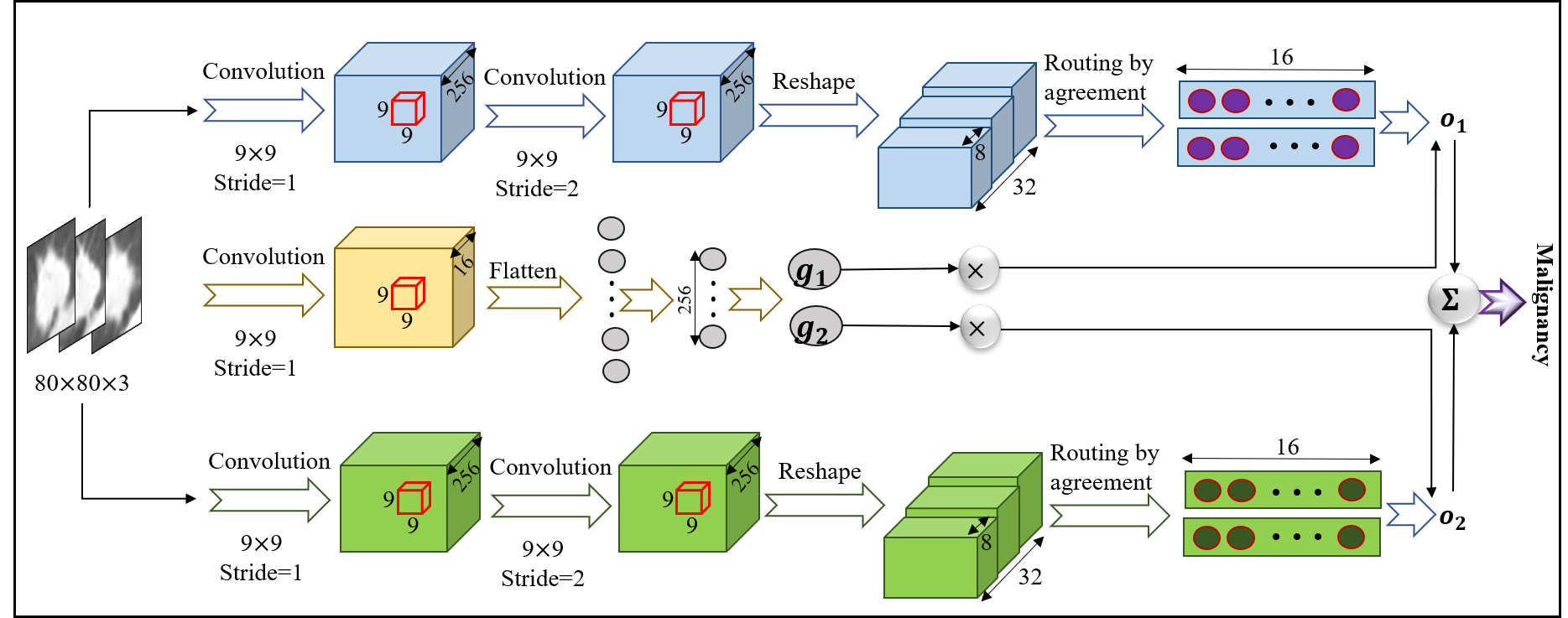}
\caption{Proposed MIXCAPS.}\label{fig:model}
\end{sidewaysfigure}

In this section, first we present the dataset used to design and develop the proposed MIXCAPS. Afterwards, the pre-processing approach, and the proposed MIXCAPS framework are described.

\subsection{Data and Pre-processing Approach}
The lung nodule malignancy dataset is adopted from the Lung Image Database Consortium (LIDC) and Image Database Resource Initiative (IDRI) dataset~\cite{Armato:2015,Armato:2011,Clark:2013}. This dataset consists of CT scans from $1,018$ subjects. All the images are labeled and annotated by one to four radiologists. Labels include non-nodule, nodule less than  $3$ $mm$ in size, and nodules with malignancy scores of $1$ to $5$, where larger numbers denote higher possibility of malignancy. In this study, we discarded all the cases with average malignancy score of $3$ which dictates an indeterminate malignancy. Consequently, we regrouped labels $1$ and $2$ as benign nodules, and labels $4$ and $5$ as malignant nodules. Therefore, we ended up having a binary classification problem with a total of $2,283$ nodules. It is worth mentioning that we included all the annotations provided by all the radiologists as separate nodules. However, the malignancy scores are the average over all the provided scores. For each nodule, we extracted a 3D patch from the center of the nodule (center slice and the two immediate neighbors). Patches are extracted to fit the nodule boundary provided by the radiologists. However, to have fixed size inputs, all patches were zero-padded to $80\times 80$ (the largest possible width and height based on the training data).

\subsection{The MIXCAPS Architecture}
The proposed capsule network-based mixture of experts for lung nodule malignancy prediction, referred to as the MIXCAPS, is shown in Fig~\ref{fig:model}. The 3D nodule patches are the inputs to two capsule network experts, as well as the convolutional gating network. The two experts, as shown in Fig~\ref{fig:model}, consist of two convolutional layers, the last of which is reshaped to form a capsule layer. This capsule layer is followed by a routing by agreement and the final capsule layer. The outputs of the two experts, denoted by $\o_1$ and $\o_2$, represent the class (benign and malignant) probabilities. The gating network, consisting of a convolutional and two fully connected layers, determines the contribution of each expert, denoted by $g_1$ and $g_2$,  for a specific input through a Softmax layer, as follows
\begin{equation}\label{eq:moe}
g_1=\frac{\exp{(G_1)}}{\exp{(G_1)}+\exp{(G_2)}}, \quad g_2=\frac{\exp{(G_2)}}{\exp{(G_1)}+\exp{(G_2)}},
\end{equation}
where $G_1$ and $G_2$ are pre-activation outputs. The Softmax layer ensures that $g_1$ and $g_2$ sum to one. These contributions are multiplied by $\o_1$ and $\o_2$ to calculate the final prediction $\o$ as follows
\begin{equation}
\o=g_1\o_1+g_2\o_2.
\end{equation}
Output vector $\o$ encompasses the probability of benign and malignant classes, denoted by $o^{(0)}$ and $o^{(1)}$, respectively. In other words
\begin{equation}
\o= [o^{(0)},o^{(1)}]^T.
\end{equation}
where superscript $T$ denotes transpose operator. Originally, margin loss is proposed for the training of the capsule networks. In this study, we adopt the same loss function with the difference that the loss $l$ is calculated over the final output of the MIXCAPS instead of the individual capsule networks, as follows
\begin{align}
&l^{(0)}=T^{(0)}\max(0,m^+-o^{(0)})^2+\lambda(1-T^{(0)})\max(0,o^{(0)}-m^-)^2,\\
&l^{(1)}=T^{(1)}\max(0,m^+-o^{(1)})^2+\lambda(1-T^{(1)})\max(0,o^{(1)}-m^-)^2,\\
&l=l^{(0)}+l^{(1)},
\end{align}
where $l^{(0)}$ and $l^{(1)}$ denote the losses associated with the benign and malignant classes, respectively. $m^+$, $\lambda$, and $m^-$ are hyper-parameters. Terms $T^{(0)}$ and $T^{(1)}$ are the ground-truth labels for benign and malignant classes, respectively. According to Reference~\cite{Jacobs:1991} comparing the desired output with the blend of outputs from the experts, leads to a strong coupling between experts and solutions in which many experts are used for one case. However, in this study, we did not encounter such a problem, and therefore did not adopt non-linear combinations of the outputs.

\subsection{CapsNet as a Mixture of Experts}
In this subsection, we revisit the idea of the capsule networks and show how they can be viewed within the mixture of experts framework. In other words, we show that a CapsNet is a series of consecutive  MoE layers such that each lower level capsule with instantiation vector $\u_i$ serves as an expert to predict the output of the capsule in the next layer  with instantiation vector $\s_j$.

Recall from Section~\ref{sec:caps} that each capsule (among $N_{PrC}$ number of primary capsules) with instantiation vector $\u_i$, for ($1 \leq i \leq N_{PrC}$),  makes predictions $\hat{\u}_{j|i}$, through Eq.~\eqref{eq:pred}. Consequently, each capsule  (among $N_{PaC}$ number of parent capsules) with  instantiation vector  $\s_j$, for ($1 \leq i \leq N_{PaC}$), receives predictions from all the lower level primary capsules. Each primary Capsule $i$, therefore, can be considered as an expert making predictions for all the parent (final) capsules. Contribution of each capsule expert $i$ to each final capsule $j$ is represented by $c_{ij}$, which is basically similar to $g_i$ in an MoE framework, with the difference that in the conventional MoE formulation, each expert contributes equally to all the outputs, whereas capsule experts have different contributions to different final capsules. This is the reason why the notation of $c_{ij}$ is used instead of $c_i$. The instantiation parameter of each final Capsule $j$ is calculated according to Eq.~\eqref{eq:sj} incorporating predictions from all the experts. Another difference between capsule experts and conventional MoE ones is that the gating model in the latter case is typically a simple or advanced machine learning model, whereas in the former case, routing by agreement serves as the gate to determine contribution through Eq.~\eqref{eq:rout} to \eqref{eq:sj}. It is also worth noting that Eq. \eqref{eq:score} ensures that contributions to each final capsule $j$ sum to one, satisfying the requirement of an MoE approach as in Eq.~\eqref{eq:moe}.

Having the aforementioned discussion in mind, each CapsNet itself is a series of mixtures of experts. In the proposed MIXCAPS, the CapsNets themselves are utilized as single experts. Therefore. MIXCAPS can be considered as a hierarchical MoE technique. It is also interesting to study how the calculation of $c_{ij}$s resembles the calculation of experts' weights in an MoE approach. Generally speaking, there are several solutions to an MoE problem~\cite{Jordan:1993}.
An Expectation Maximization (EM) algorithm is one applicable solution, through which the experts' weights are considered as hidden variables, whose posteriors are estimated in the E-step, as follows
\begin{equation}
p(z^n_i|\t^n,\x^n)=\frac{p(\t^n|z^n_i=1,\x^n)p(z^n_i=1|\x^n)}{p(\t^n|\x^n)},\label{eq:e1}
\end{equation}
where binary variable $z^n_i$ is one when instance $n$ is assigned to expert $i$, and zero otherwise. Term $p(z^n_i|\t^n,\x^n)$ represents the posterior probability of $z^n_i$ given input vector $\x^n$ and target vector $\t^n$. Following the Bayes' rule, this posterior is calculated using the likelihood term $p(\t^n|z^n_i=1,\x^n)$ and the prior over $z^n_i$, denoted by $p(z^n_i=1|\x^n)$. All the terms appearing in Eq.~\eqref{eq:e1} can be calculated through the MIXCAPS framework. The likelihood term can be replaced by the output of the expert capsule networks $o_i^{n(1)}$, which denotes the probability of malignancy for Instance $n$, based on the $i^{th}$ expert. The prior probability can also be estimated using the output of the gating model $g_i^n$ denoting the probability of assigning Instance $n$ to Expert $i$. The posterior, therefore, can be defined as
\begin{equation}
p(z^n_i|\t^n,\x^n)=\frac{g_i^no_i^{n(1)}}{\sum_{j}^{M}g_j^no_j^{n(1)}},\label{eq:e}
\end{equation}
where $M$ is the number of experts. 

To further shed light on the MoE view of CapsNets, it would be interesting to note that the EM formulation of the MoE  closely resembles the weight update process of a multiple model (MM)~\cite{Mohammadi:2015} approach. In MM formulation, observations are sequentially generated from different models and the goal is to identify the contribution of each single model $i$ given all the observations up to the current time ($\Y^k$), as follows
\begin{equation}
p(z^k_i|\Y^k)=\frac{p(\y^k|z^k_i=1,\Y^{k-1})p(z^k_i=1|\Y^{k-1})}{\sum_{j=1}^{M}p(\y^k|z^k_j=1,\Y^{k-1})p(z^k_j=1|\Y^{k-1})},\label{eq:mm}
\end{equation}
where $\y^k$ is the most recent observation. Comparing Eq.~\eqref{eq:mm} with Eq.~\eqref{eq:e}, it can be seen that while the prior in an MoE approach is determined based on the current input vector, it is calculated based on the previous observations, in the MM case. In other words, in MM, the prior is iteratively replaced with the posterior. The updates of coefficients in the routing by agreement process of the CapsNet is similar to the weight updates in MM. In particular, in each round of the routing by agreement, the previously calculated $c_{ij}$ serves as the prior to compute the coefficient in the next round.

\section{Results and Discussion}\label{sec:Result}
\begin{sidewaystable}
\centering
\caption{Performance of the proposed MIXCAPS compared to that of a single capsule network and a mixture of CNNs. Numbers in parenthesis show the 95\% confidence intervals.\label{tab:result}}
\vspace{.05in}
\begin{tabular}{|l|l|l|l|l|}
\hline
\textbf{Model} & \textbf{Sensitivity} & \textbf{Specificity} & \textbf{Accuracy} & \textbf{AUC} \\
\hline
\textbf{Proposed MIXCAPS} & $\textbf{89.5} (89.3, 89.7) \%$ & $\textbf{93.4} (93.2, 93.6) \%$ & $\textbf{90.7} (90.6, 90.8) \%$ & $\textbf{0.956} (0.955, 0.956)$\\
\hline
\textbf{Single capsule network} & $ 86.1 (85.7, 86.4) \%$ & $90.8 (90.5, 91.1) \%$ & $88.6 (88.5, 88.7) \%$ & $0.938 (0.937, 0.939)$\\
\hline
\textbf{Mixture of CNNs} & $87.5 (87.1, 87.8) \%$ & $91.3 (91.1, 91.6) \%$ & $89.5 (89.4, 89.7) \%$ & $0.948 (0.946, 0.948) $\\
\hline
\end{tabular}
\end{sidewaystable}
\begin{sidewaystable}
\centering
\caption{\label{tab:papers}List of studies that have used LIDC-IDRI to predict lung nodule malignancy based on the ratings provided by radiologists. Note that some of the studies  such as References~\cite{Causey:2018} and~\cite{Xie:2017} included hand-crafted features, requiring expert annotations. Numbers in parenthesis show the 95\% confidence intervals obtained from 200 iterations of bootstrapping.}
\begin{tabular}{|l|l|l|l|l|}
\hline
\textbf{Method}  & \textbf{Area Under the Curve (AUC)} & \textbf{Accuracy} & \textbf{Specificity} & \textbf{Sensitivity} \\
\hline
Proposed MIXCAPS & $0.956 (0.955, 0.956)$ & $90.7 (90.6, 90.8) \%$ & $89.5 (89.3, 89.7) \%$ & $89.5 (89.3, 89.7)$\\
\hline
CNN~\cite{Causey:2018} & $0.938$ & $87.9\%$ & $87.9\%$ & $87.9\%$\\
\hline
CNN in combination with hand-crafted features~\cite{Causey:2018} & $\textbf{0.971}$ & $\textbf{93.2}\%$ & $\textbf{98.5}\%$ & $87.9\%$\\
\hline
Deep residual network~\cite{Nibali:2017} & $0.9459$ & $89.90\%$ & $88.64\%$ & $\textbf{91.07}\%$\\
\hline
Deep belief network~\cite{Sun:2016} & - & $81.19\%$ & - & -\\
\hline
CNN in combination with hand-crafted features~\cite{Xie:2017} & - & $86.79\%$ & $95.42\%$ & $60.26\%$\\
\hline
Multi-crop CNN~\cite{Shen:2017} &$0.93$ & $87.14\%$ & $93\%$ & $77\%$\\
\hline
\end{tabular}
\end{sidewaystable}

In this section, three different experiments on lung cancer malignancy prediction are presented. The main objective is to evaluate performance of the proposed MIXCAPS framework and compare its capabilities with those of its state-of-the-art counterparts. Results are obtained with 200 iterations of bootstrapping, where in each iteration, 80\% of the data is sampled (with replacement) from the whole dataset. 20 \% of the training dataset is then randomly extracted for validation. A 95\% confidence interval (CI) is calculated for all the performance metrics. 
Adam optimizer with $10$ epochs and batch size of $16$ is used for training.

\vspace{.1in}
\noindent
\textbf{\textit{Experiment 1}}:
Our first experiment is to compare the performance of the proposed MIXCAPS with a single capsule network and a mixture of CNNs, as shown in Table~\ref{tab:result}, where performance is measured in terms of sensitivity, specificity, accuracy, and area under the curve (AUC). The architecture of the single capsule network is exactly the same as the CapsNet experts. We tried to keep the complexity as similar as possible to the MIXCAPS, when designing the mixture of CNNs. In particular, the gating network exactly resembles that of the MIXCAPS. The CNN experts consist of two convolutional layers with $256$ filters, similar to the experts in the MIXCAPS. The convolutional layers are followed by a dense layer with 32 neurons (the same as the dimension of the last capsule layers), and the final softmax layer for nodule malignancy prediction. As shown in Table~\ref{tab:result}, MIXCAPS outperforms its two aforementioned counterparts, in terms of sensitivity, specificity, accuracy, and AUC. 

\vspace{.1in}
\noindent
\textbf{\textit{Experiment 2}}:
In the second experiment, we compare the proposed MIXCAPS with several well-known studies on the same dataset. Table~\ref{tab:papers} shows these studies, their methods, and the obtained results. As it can be inferred from Table~\ref{tab:papers}, the proposed MIXCAPS outperforms all the studies in terms of accuracy and AUC, except Reference~\cite{Causey:2018}. However, it is worth mentioning that the aforementioned study utilizes hand-crafted radiomics, requiring fine annotation of the nodules, from which our proposed approach is independent. Reference~\cite{Xie:2017} has obtained a higher specificity compared to the proposed MIXCAPS. Its low sensitivity, however, is a sign of an unbalanced classification and/or over-fitting. Reference~\cite{Nibali:2017} has achieved the highest sensitivity among all the other references. Nevertheless, no confidence interval is provided to ensure the robustness of the result.

\begin{figure}[th]
\centering
\includegraphics[scale=.5]{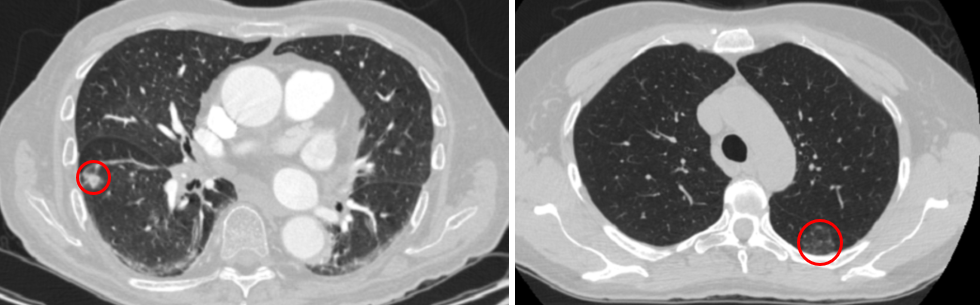}
\caption{Example of nodules assigned to experts based on  their volume and diameter. The nodule on the left, which has a lower probability of belonging to the first expert, is smaller in terms of volume and diameter compared to the nodule on the right.}\label{fig:data}
\end{figure}

\vspace{.1in}
\noindent
\textbf{\textit{Experiment 3}}:
Finally, we conduct an experiment to gain an insight on how the data instances are split between the two experts. The LIDC-IDRI dataset is accompanied by a few nodule-related properties, determined by the radiologists. These features include volume, diameter, $x$ center of mass and $y$ center of mass. We calculated the correlation between the output of the gating network and these features. While the correlations with volume and diameter are $0.58$ and $0.77$, respectively, we observed no correlation with the centers of mass. It should be noted that the inputs to the proposed MIXCAPS are cropped nodule regions. In other words, the model has no access to the location of the nodule. Therefore, the almost zero correlations with the centers of the mass is completely expected. The observed correlations between the gate outputs and the volume and diameter imply that larger nodules have higher probabilities of being assigned to the first expert. Fig.~\ref{fig:data} shows two nodules in the test set. The left nodule, which has a volume of $496.32$ and diameter of $9.823$, has a low probability of belonging to the first expert, whereas the nodule on the right, with a volume of $6663.44$ and diameter of $23.347$, has a high probability of being assigned to the first expert. In other words, the first expert tends to handle larger nodules, compared to the second expert.

Although MoE techniques are shown to be able to improve the classification performance, they typically face an objection related to the high computational cost at the test time. This problem, however, can be dealt with by using distillation~\cite{Hinton:2015}. Therefore, in our future studies, we will focus on distilling MIXCAPS into a smaller and more time-efficient model.

\subsection{MIXCAPS for Brain Tumor Type Classification}
\begin{sidewaysfigure}
\centering
\includegraphics[scale=.6]{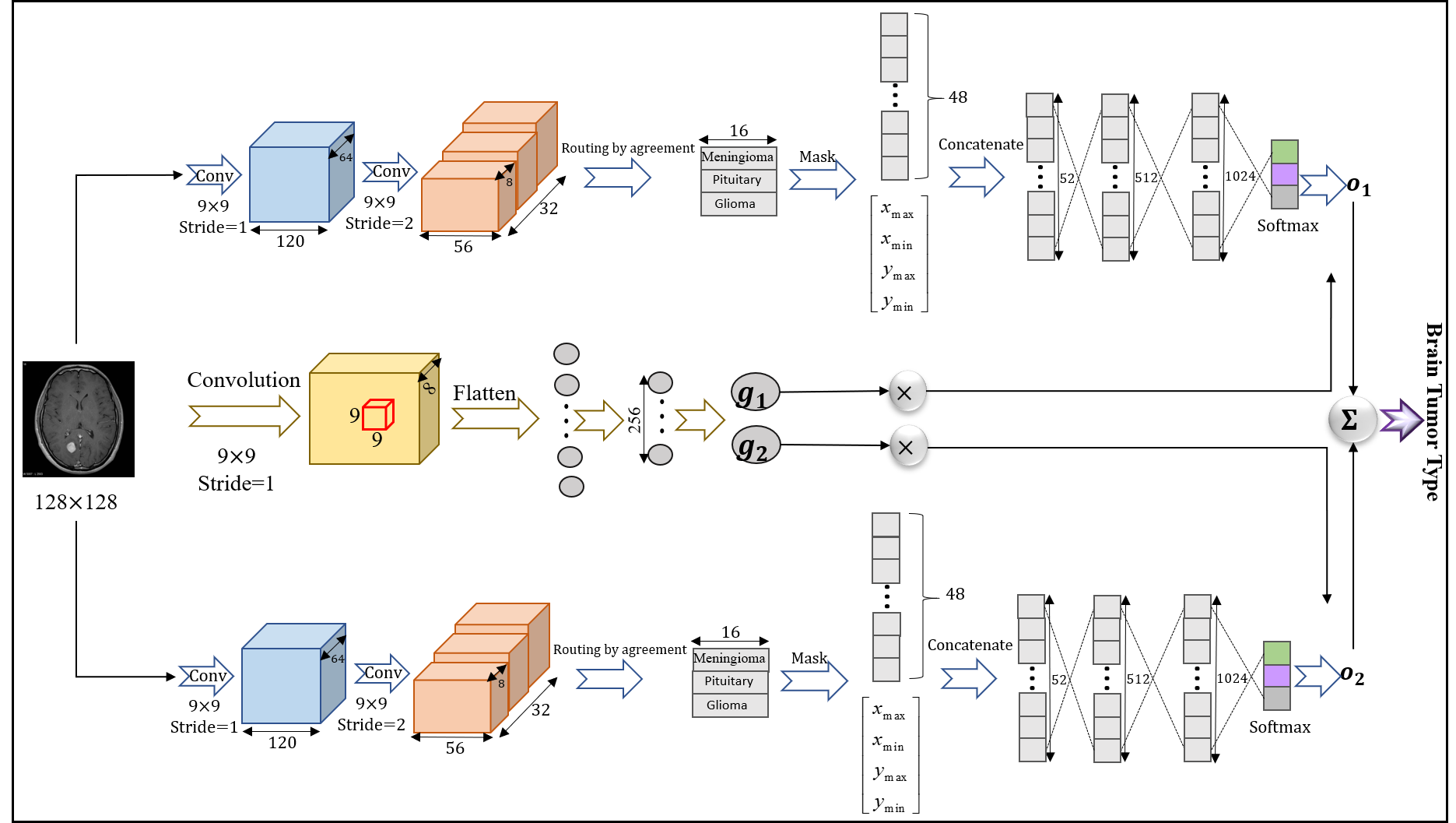}
\caption{MIXCAPS architecture with BoxCaps as experts for brain tumor type classification.}\label{fig:brainmodel}
\end{sidewaysfigure}
Brain tumor is among the deadliest cancers. Determining the type of the tumor, which is a challenging task in terms of accuracy and inter-observer variability, can significantly facilitate the control/treatment process. Therefore, we dedicate this subsection to investigate the generalizability of the proposed MIXCAPS to brain tumor type classification. In a previous study~\cite{Afshar:2019ICASSP}, we proposed a capsule network-based framework, which we referred to as the BoxCaps, for brain tumor classification, considering not only raw magnetic resonance imaging (MRI) inputs, but also the coarse tumor boundaries. The motivation behind such framework was that the whole brain image contained valuable information on the location of the tumor with respect to the brain tissue. The CapsNet, however, tends to get distracted from the main tumor region when being fed with all the details from the brain image. As such, we designed a modified architecture where the output capsules were concatenated with the tumor course boundary box. This way, the model had access to both brain tissue and tumor region.

To investigate whether the MIXCAPS can be generalized to brain tumor classification, we replaced the capsule experts in MIXCAPS with the previously designed BoxCaps architecture, as shown in Fig.~\ref{fig:brainmodel}. We then tested the resulting framework on a brain tumor dataset~\cite{Cheng:2016}, where train, validation, and test splits are obtained from the same bootstrapping approach used for the LIDC-IDRI dataset. The aforementioned dataset consists of $3,064$ images from $233$ patients, diagnosed with one of the three brain tumor types, i.e., Meningioma, Pituitary, and Glioma. Table~\ref{tab:resultbrain} presents the obtained results, according to which, the MoE approach leads to higher accuracy compared to a single BoxCaps. Furthermore, the MoE approach leads to higher sensitivity for Glioma and Pituitary, and higher specificity for Meningioma and pituitary tumor types.

\begin{table*}[ht]
\centering
\caption{Performance of the proposed MIXCAPS with BoxCaps as experts. Numbers in parenthesis show the 95\% confidence intervals. \label{tab:resultbrain}}
 \begin{tabular}{|c| c| c|}
 \cline{2-3}
 \multicolumn{1}{c|}{} & \textbf{MIXCAPS-BoxCaps} & \textbf{BoxCaps}\\
 \hline
 \rowcolor{c1}
 Accuracy & \textbf{91.3} (91.1, 91.5) \% & 90.9 (90.2, 91.5) \% \\
 \hline
  \rowcolor{c2}
 Sensitivity for Meningioma & 77.5 (77.1, 77.9) \% & \textbf{80.1} (76.2, 84) \% \\
 \hline
 \rowcolor{c2}
 Sensitivity for Glioma & \textbf{95.9} (93.2, 98.5) \% & 92 (90, 94.1) \% \\
 \hline
 \rowcolor{c2}
 Sensitivity for Pituitary  & \textbf{97.7} (97.2, 98.3) \% & 97.2 (95.6, 98.9) \% \\
 \hline
 \rowcolor{c3}
 Specificity for Meningioma & \textbf{96.1} (96, 96.1) \% & 94.1 (92.7, 95.5) \% \\
 \hline
 \rowcolor{c3}
 Specificity for Glioma & 88.7 (87.6, 89.8) \% & \textbf{89.8} (88.4, 91.2) \% \\
 \hline
 \rowcolor{c3}
 Specificity for Pituitary  & \textbf{88.7} (86.2, 91.2) \% & 88.1 (86.9, 89.3) \% \\
 \hline
 \end{tabular}
\end{table*}

Finally, we conduct another experiment to study if the provided boundary box is the only important factor leading to the obtained result. In other words we need to make sure that the input images are not ignored by the model, simply because the boundary box itself can determine the tumor type. To this end, we gradually added zero-mean Gaussian noise to input images and calculated the model's accuracy. It is observed that while a noise with a standard deviation (STD) of $0.01$ does not change the accuracy, increasing STD to $0.1$ and $0.5$ degrades the accuracy to $84.44\%$ and $76\%$, respectively. This experiment shows that while the boundary box assists the classification, it does not replace the input images.

\section{Conclusion and Future Direction}\label{Sec:Conclusion}
In this paper, we proposed a capsule network-based mixture of experts framework, referred to as the MIXCAPS, for lung nodule malignancy prediction. The proposed MIXCAPS frameworks contains two capsule network experts and a convolutional gating network to assign instances to experts. Our obtained results show that MIXCAPS outperforms a single capsule network and a mixture of CNNs. It has also several advantages over the previous methods. First, MIXCAPS utilizes capsule networks and is therefore capable of handling smaller datasets. Second, through the MoE approach, experts get the chance to specialize on a subset of the data. Furthermore, MIXCAPS does not require fine annotations and is independent from pre-defined hand-crafted features. Our future directions include exploring capsule gating networks and optimizing the number of experts, as well as focusing on MIXCAPS knowledge distillation to improve the model's time-efficacy.

\bibliography{refs}

\end{document}